\documentclass[useAMS,usenatbibi,usegraphicx]{mn2e}
\usepackage{psfig}

\newcommand\beq{\begin{equation}}
\newcommand\eeq{\end{equation}}

\newcommand{\be}{\begin{equation}}
\newcommand{\ee}{\end{equation}}
\newcommand{\ba}{\begin{eqnarray}}
\newcommand{\ea}{\end{eqnarray}}

\newcommand{\Ms}{M_{\odot}}

\newcommand{\bml}{\begin{mathletters}}
\newcommand{\eml}{\end{mathletters}}

\def\ltsima{$\; \buildrel < \over \sim \;$}
\def\simlt{\lower.5ex\hbox{\ltsima}}
\def\gtsima{$\; \buildrel > \over \sim \;$}
\def\simgt{\lower.5ex\hbox{\gtsima}}
\def\gsim{ \lower .75ex \hbox{$\sim$} \llap{\raise .27ex \hbox{$>$}} }
\def\lsim{ \lower .75ex\hbox{$\sim$} \llap{\raise .27ex \hbox{$<$}} }
\def\msun{\,{\rm M_\odot}}

\title[Massive black holes and pulsar timing]
{Gravitational waves from resolvable massive black hole binary systems and observations with Pulsar Timing Arrays}

\author[A. Sesana et al.]{A. Sesana$^{1}$, A. Vecchio$^{2}$ and M. Volonteri$^3$\\
$^{1}$Center for Gravitational Wave Physics, The Pennsylvania State University, University Park, PA 16802, USA\\
$^{2}$School of Physics and Astronomy, University of Birmingham, 
Edgbaston, Birmingham, B15 2TT, UK\\
$^{3}$ Dept. of Astronomy, University of Michigan, Ann Arbor, MI 48109, USA}

\begin{document}

\date{Received ---}

\maketitle
\begin{abstract}
Massive black holes are key components of the assembly and evolution of cosmic structures and a number of surveys are currently on-going or planned to probe the demographics of these objects and to gain insight into the relevant physical processes. Pulsar Timing Arrays (PTAs) currently provide the only means to observe gravitational radiation from massive black hole binary systems with masses $\simgt 10^7\Ms$. The whole cosmic population produces a stochastic background that could be detectable with upcoming Pulsar Timing Arrays. Sources sufficiently close and/or massive generate gravitational radiation that significantly exceeds the level of the background and could be individually resolved. We consider a wide range of massive black hole binary assembly scenarios, we investigate the distribution of the main physical parameters of the sources, such as masses and redshift, and explore the consequences for Pulsar Timing Arrays observations. Depending on the specific massive black hole population model, we estimate that on average at least one resolvable source produces  timing residuals in the range $\sim5-50$ ns. Pulsar Timing Arrays, and in particular the future Square Kilometre Array (SKA), can plausibly detect these unique systems, although the events are likely to be rare. These observations would naturally complement on the high-mass end of the massive black hole distribution function future surveys carried out by the Laser Interferometer Space Antenna ({\it LISA}).
\end{abstract}

\section{introduction}

Massive black hole (MBH) binary systems with masses in the range
$\sim 10^4-10^{10}\msun$ are amongst the primary candidate 
sources of gravitational waves (GWs) at $\sim$ nHz - mHz frequencies (see, e.g., 
Haehnelt 1994; Jaffe \& Backer 2003; Wyithe \& Loeb 2003, Sesana et al. 2004, 
Sesana et al. 2005). The frequency band $\sim 10^{-5}\,\mathrm{Hz} - 1 \,\mathrm{Hz}$ 
will be probed by the {\it Laser Interferometer Space Antenna} ({\it LISA}, Bender et al. 1998),
a space-borne gravitational wave laser interferometer being developed by ESA and NASA. 
The observational window $10^{-9}\,\mathrm{Hz} - 10^{-6} \,\mathrm{Hz}$ is already accessible 
with Pulsar Timing Arrays (PTAs;  e.g. the Parkes radio-telescope, Manchester 2008). PTAs exploit
the effect of GWs on the propagation of radio signals from a pulsar to the Earth (e.g. Sazhin 1978, Detweiler 1979, Bertotti et al. 1983), producing a characteristic signature in the time of arrival (TOA) of radio pulses.  The timing residuals of the fit of the actual TOA of the pulses and the TOA according to a given model, carry the physical information about unmodelled effects, including GWs (e.g. Helling \& Downs 1983, Jenet et al. 2005). The complete Parkes PTA (Manchester 2008), the European Pulsar Timing Array (Janssen et al. 2008), and NanoGrav \footnote{http://arecibo.cac.cornell.edu/arecibo-staging/nanograv/} are expected to improve considerably on the capabilities of these surveys and the planned Square Kilometer Array (SKA; {\it www.skatelescope.org}) will produce a major leap in sensitivity. 

Popular scenarios of MBH formation and evolution (e.g. Volonteri, Haardt \& Madau 2003; Wyithe \& Loeb 2003, Koushiappas \& Zentner 2006, Malbon et al. 2007, Yoo et al. 2007) predict the existence of a large number of massive black hole binaries (MBHB) emitting in the frequency range between $\sim 10^{-9}$ Hz and $10^{-6}$ Hz. PTAs can gain direct access to this population, and address a number of unanswered questions in astrophysics (such as the assembly of galaxies and dynamical processes in galactic nuclei), by detecting gravitational radiation of two forms: (i) the stochastic GW background produced by the incoherent superposition of radiation from the whole cosmic population of MBHBs and (ii) GWs from individual sources that are sufficiently bright (and therefore massive and/or close) so that the gravitational signal stands above the root-mean-square (rms) value of the background. Both classes of signals are of great interest, and the focused effort on PTAs could lead to the discovery of systems difficult to detect with other techniques.

The possible level of the GW background, and the consequences for observations have been explored by several authors (see {\em e.g.} Rajagopal \& Romani 1995; Phinney 2001, Jaffe \& Backer 2003; Jenet et al. 2005; Jenet et al. 2006; Sesana et al. 2008). Recently, Sesana Vecchio \& Colacino (2008, hereinafter PaperI) studied in details the properties of such a signal and the astrophysical information encoded into it, for a comprehensive range of MBHB formation models. As shown in PaperI, there is over a factor of 10 uncertainty in the characteristic amplitude of the MBHB generated background in the PTA frequency window. However, the most optimistic estimates yield an amplitude just a factor $\approx 3$ below the upper-bound placed using current data (Jenet et al. 2006), and near-term future observations could either detect such a stochastic signal or start ruling out selected MBHB population scenarios. Based on our current astrophysical understanding of the formation and evolution of MBHBs and the estimates of the sensitivity of SKA, one could argue that this instrument guarantees the detection of this signal in the frequency range $3\times 10^{-9}\,\mathrm{Hz} - 5\times 10^{-8} \,\mathrm{Hz}$ for essentially every assembly scenario that is considered at present. 

The background generated by the cosmic population of MBHBs is present across the whole observational window of PTAs (cf. PaperI). The Monte Carlo simulations reported in PaperI show clearly the presence of distinctive strong peaks well above the average level of the stochastic contribution (cf. Figure 1 and 4 in PaperI). This is to be expected, as individual sources can generate gravitational radiation sufficiently strong to stand above the rms value of the stochastic background. These sources are of great interest because they can be individually resolved and likely involve the most massive MBHBs in the Universe. Their observation can offer further insight into the high-mass end of the MBH(B) population, galaxy mergers in the low-redshift Universe and dynamical processes that determine the formation of MBH pairs and the evolution to form close binaries with orbital periods of the order of years. 

Some exploratory studies have been carried out about detecting individual signals from MBHBs in PTA data (Jenet et al. 2004, 2005). In this paper we study systematically for a comprehensive range of assembly scenarios the properties, in particular the distribution of masses and redshift, of the sources that give rise to detectable individual events; we compute the induced timing residuals and the expected number of sources at a given timing residual level.
To this aim, the modelling of the high-mass end of the MBHB population at relatively low redshift is of crucial importance. We generate a statistically significant sample of merging massive galaxies from the on-line Millennium database  ({\it http://www.g-vo.org/Millennium}) and populate them with central MBHs according to different prescriptions (Tremaine et al. 2002, Mclure et al. 2006, Lauer et al. 2007, Tundo et al. 2007). The Millennium simulation (Springel et al. 2005) covers a comoving volume of $(500/h_{100})^3$ Mpc$^3$ ($h_{100}=H_0/100$  km s$^{-1}$ Mpc$^{-1}$ is the normalized Hubble parameter), ensuring a number of massive nearby binaries adequate to construct the necessary distribution. For each model we compute the stochastic background, the expected distribution of bright individual sources and the value of the characteristic timing residual $\delta t_\mathrm{gw}$, see Equation (\ref{e:deltatgw}), for an observation time $T$. The signal-to-noise ratio at which a source can then be observed scales as SNR$\approx \delta t_\mathrm{gw}/\delta t_\mathrm{rms}$ where $\delta t_\mathrm{rms}$ is the root-mean-square level of the timing residuals noise, both coming from the receiver {\em and} the GW stochastic background contribution. In the following we summarise our main results:
\begin{enumerate} 
\item The number of detectable individual sources for different thresholds of the effective induced timing residuals $\delta t_\mathrm{gw}$ is shown in Table \ref{tab:summary}. Depending on the specific MBH population model, we estimate that on average at least one resolvable source produces  timing residuals in the range $\sim5-50$ns. Future PTAs, and in particular SKA, can plausibly detect these unique systems; the detection is however by no means guaranteed, events will be rare and just above the detection threshold.

\item As expected, the brightest signals come from very massive systems with
${\cal M}>5\times10^8\msun$. Here ${\cal M}=M_1^{3/5}M_2^{3/5}/(M_1+M_2)^{1/5}$ is the chirp mass of the binary and $M_1>M_2$ are the two black hole masses. Most of the resolvable sources are located at relatively high redshift ($0.2 < z < 1.5$), and not at $z \ll 1$ as one would naively expect, giving the opportunity to probe the universe at cosmological distances.

\item The number of resolvable MBHBs depends on the actual level of the stochastic background generated by the whole population; here we have used the standard simplified assumption that the background level is determined by having more than one source per frequency resolution bin of width $1/T$, where $T$ is the observational time. Using this 
definition we find that at frequencies less than $10^{-7}$ Hz there are typically a few
resolvable sources, considering $T= $ 5 yrs, with residuals in the range $\sim 1\,\mathrm{nHz} - 1\,\mu\mathrm{Hz}$.
As the level of the background decreases for increasing frequencies, fainter sources become visible individually.

\item As a sanity check, we have compared the MBHB populations and stochastic background levels obtained using data from the Millennium simulation (adopted in this paper) with those derived by means of merger tree realisations based on the Extended Press \& Schechter  (EPS) formalism (considered in PaperI) and have found good agreement. This provides an additional validation of the results of this paper and PaperI. Moreover it supports that EPS merger trees, if handled sensibly, can offer a valuable tool for the study of MBH evolution even at low redshift.
\end{enumerate}
  
The paper is organised as follows. In Section 2 we describe MBHB population models, in particular the range of scenarios considered in this paper. A short review of the timing residuals produced by GWs generated by an individual binary (in circular orbit) in the data collected by PTAs is provided in Section 3. 
Section 4 contains the key results of the paper: the expected timing residuals from the estimated population of MBHBs, including detection rates for current and future PTAs. We also provide a comparison between the stochastic background computed according to the prescriptions considered here and the results of PaperI. Summary and conclusions are given in Section 5. 

\section{The massive black hole binary population}

In this section we introduce the population synthesis model adopted to estimate the number, and astrophysical parameters of MBHBs that emit GWs in the frequency region probed by PTAs. The two fundamental ingredients to compute the merger rate of MBHBs are (i) the merger history of galaxy haloes, and (ii) the MBH population associated to those haloes. We discuss them in turn. In building and evolving the population of sources we follow exactly the same method as in PaperI, to which we refer the reader for further details, with one important difference: the galactic haloes merger rates are derived using the data provided by the Millennium simulation, and not EPS-based models. We will justify this choice in the next sub-Section, but note that the two methods yield (within the statistical error) the same results. This is a result that is important in itself and has far reaching consequences (outside the specific issues related to PTAs). 

\subsection{From merger trees to the Millennium database}

In PaperI we used models based on the Extended Press \& Schechter  (EPS) formalism (Press \& Schechter 1974, Lacey \& Cole 1993, Sheth \& Tormen 1999), that trace the hierarchical assembly of dark matter haloes through a Monte-Carlo approach. Although EPS--based models tend to overpredict the bright end of the quasar luminosity function at $z<1$ (e.g. Marulli et al. 2006), we showed that EPS halo merger rates at low redshift are consistent with observations of close galaxy pairs (Lin et al. 2004, Bell et al. 2006, De Propris et al. 2007). In this paper we focus on MBHBs, whose GWs induce timing residuals above the stochastic signal from the whole population, and are therefore detectable as individual sources. The population of low/medium-redshift and high-mass sources will particularly impact on the results. At low  redshift, the EPS-based merger tree outputs need to be handled with care. In models such as those considered in PaperI, each realization of the Universe is obtained by reconstructing the merger history of about 200 dark matter haloes, see Volonteri et al. (2003) for details. The outputs of the models are a list of coalescences labelled by MBH masses (for a given recipe that associates a MBH mass and to a given dark matter halo) and redshift. These events are then properly weighted over the observable volume shell at each redshift to obtain the distribution $d^3N/d{\cal M}dzdt$ (see PaperI), that is the coalescence rate (the number of coalescences $N$ per time interval $dt$) in the chirp mass and redshift interval $[{\cal M},{\cal M}+d{\cal M}]$ and $[z, z+dz]$, respectively. The resulting distribution is reasonably smooth over most of the (${\cal M},z$) plane, but small number statistics becomes important at $z<0.5$ and ${\cal M}>10^9\msun$, which is an important region of the parameter space when one deals with individual sources.

To avoid this problem, in this paper we generate distributions of coalescing MBHBs using the galaxy haloes merger rates derived  from the on-line Millennium run database. The Millennium simulation  
(Springel et al. 2005) covers a volume of $(500/h_{100})^3$ Mpc$^3$ and is the ideal tool to construct a statistically representative distribution of massive low/medium--redshift sources. In fact, the typical 
ensemble of events available to construct the mass function of coalescing binaries
is $\sim100$ times larger than in a typical EPS-based merger tree realization. As a first step we compile 
catalogues of galaxy mergers from the semi-analytical model of Bertone et al. (2007) applied to the 
Millennium run.

\subsection{Populating galaxies with massive black holes}
We need to associate to each merging galaxy in our catalogue 
a central MBH, according to some sensible prescription.
The Bertone et al. 2007 catalogue contains many properties
of the merging galaxies, including the bulge mass $M_{\rm bulge}$, and the 
bulge rest frame magnitude $M_V$ both of the progenitors and of the merger remnant.
This is all we need in order to populate a galaxy with a central MBH.  
The process is twofold.

\begin{table*}
\begin{center}
\begin{tabular}{ll|cccc}
\hline
        				&					& $M_{\rm BH}-M_{\rm bulge}$ 	& $M_{\rm BH}-M_{\rm bulge}$	& $M_{\rm BH}-M_V$	& $M_{\rm BH}-\sigma$ 	\\
				&					& Tundo et (2007)				&	Mclure et al. (2006)				&  Lauer et al. (2007)	& Tremaine et al. (2002)	\\
\hline
Single BH accretion	        & 	 				& Tu-SA		 & Mc-SA		& La-SA		 & Tr-SA	\\
 	&					                        &$6.9\,(2.7)$    &$6.6\,(2.5)$	        &$8.1\,(3.0)$    &$6.2\,(2.5)$	\\
				&                                       &$1.5\,(1.1)$	 &$1.0\,(0.8)$	        &$1.7\,(1.2)$	 &$0.8\,(0.8)$	\\
				&					&$0.2\,(0.4)$	 &$0.02\,(0.1)$  	&$0.5\,(0.6)$	 &$0.01\,(0.1)$	\\
				&					&$0.04\,(0.2)$	 &$0.002\,(0.04)$	&$0.1\,(0.2)$	 &$0.002\,(0.04)$	\\
				&					&	 &	&	 &	\\
Double BH accretion 	        &					& Tu-DA	         & Mc-DA		& La-DA		 & Tr-DA	\\
	&					                        &$8.3\,(2.9)$    &$7.3\,(2.7)$	&$9.6\,(3.2)$    &$7.0\,(2.8)$	\\
				&                                       &$2.2\,(1.4)$	 &$1.6\,(1.1)$	&$2.6\,(1.5)$	 &$1.2\,(1.0)$	\\
				&					&$0.6\,(0.7)$	 &$0.2\,(0.4)$	&$0.8\,(0.8)$	 &$0.06\,(0.2)$	\\
				&					&$0.2\,(0.4)$	 &$0.03\,(0.2)$	&$0.3\,(0.5)$	 &$0.007\,(0.1)$	\\
				&					&	 &	&	 &	\\
No accretion (before merger)    &					& Tu-NA		 & Mc-NA		& La-NA		 & Tr-NA	        \\
				&					&$6.4\,(2.5)$	 &$6.0\,(2.4)$	  &$6.8\,(2.7)$          &$6.0\,(2.5)$	\\
    		        	        &                               &$1.3\,(1.0)$	 &$0.5\,(0.6)$	  &$1.5\,(1.1)$	         &$0.5\,(0.6)$	\\
				&					&$0.1\,(0.3)$	 &$0.07\,(0.1)$	  &$0.1\,(0.3)$	         &$0.003\,(0.05)$	\\
				&					&$0.02\,(0.1)$	 &$0.001\,(0.03)$ &$0.02\,(0.1)$	 &$-\,(-)$	\\
\hline				
\end{tabular}
\end{center}
\caption{The table summarises the 12 models of assembly of massive black hole binary populations considered in the paper -- "Tu", "Mc", "La" and "Tr" identify the MBH-host relation; ``SA", ``DA'', and ``NA'' label the accretion mode; full details on the models are given in Section 2 -- and the total number of individually resolvable systems $N(\delta t_\mathrm{gw})$ for selected values of the characteristic timing residuals (for each model, from top to bottom $\delta t_\mathrm{gw} = 1, 10, 50$ and 100 ns considering an integration time of 5 yrs). The values in the table are the sample mean and standard deviation within brackets computed over the 1000 Monte-Carlo realisations for each model. }
\label{tab:summary}
\end{table*}

\begin{enumerate}
\item We populate the coalescing galaxies with
MBHs according to four different MBH-host prescriptions:
\begin{itemize}
\item $M_{\rm BH}-M_{\rm bulge}$ in the version given by Tundo et al. (2007, ``Tu'' models, see Table 1):
\begin{equation}
\frac{M_{\rm BH}}{\msun}=8.31+1.12\,{\rm log}\left(\frac{M_{\rm bulge}}{10^{11}\msun}\right),
\label{tundo}
\end{equation}   
with a dispersion $\Delta=0.33$ dex.

\item $M_{\rm BH}-M_{\rm bulge}$, with a redshift dependence in the version given by
Mclure et al. (2006, ``Mc'' models, see Table 1):
\begin{equation}
\frac{M_{\rm BH}}{M_{\rm bulge}}=2.07\,{\rm log}(1+z) -3.09,
\label{mclure}
\end{equation}   
with a redshift dependent dispersion $\Delta=0.125z+0.25$ dex
(see Figure 3 of Mclure et al., 2006).

\item $M_{\rm BH}-M_V$ as given by Lauer et al. (2007, ``La'' models, see Table 1):
\begin{equation}
\frac{M_{\rm BH}}{\msun}=8.67-1.32\left(\frac{M_V+22}{2.5}\right),
\end{equation}   
\label{lauer}
with dispersion $\Delta=0.35$ dex.

\item $M_{\rm BH}-\sigma$ as given by Tremaine et al. (2002, ``Tr'' models, see Table 1):
\begin{equation}
\frac{M_{\rm BH}}{\msun}=8.13+4.02\,{\rm log}{\sigma_{200}},
\label{tremaine}
\end{equation}   
 where $\sigma_{200}$ is the velocity dispersion in units of $200$ km s$^{-1}$, and 
the assumed dispersion of the relation is $\Delta=0.3$ dex. $\sigma$ is
obtained applying the Faber-Jackson (Faber \& Jackson 1976) relation in the form reported by
Lauer et al. 2007 to the values of $M_V$ obtained by the catalogue.
\end{itemize}
To each merging system we assign MBH masses according to equations
(\ref{tundo})-(\ref{tremaine}) so that we have the masses of 
the two MBH progenitors, $M_1$ and $M_2$. For each prescription we also calculate
the mass of the MBH remnant, $M_r$, using the same equations.  
In all cases (\ref{tundo})-(\ref{tremaine}), the remnant mass is $M_r>M_1+M_2$, 
consistent with the fact that MBHs are expected to grow predominantly by accretion. We also emphasize that the observed scatter is included in each relation, 
according to the observational evidence that similar bulges may host significantly different MBHs.

\item For each MBH-host relation we consider three different accretion scenarios:
\begin{itemize}
\item The masses of the coalescing MBHs are $M_1$ and $M_2$. That is, either no accretion occurs, and the merger remnant, $M_1+M_2<M_r$, sits below the predicted mass, or accretion is triggered {\it after} the MBHB coalescence. We label this accretion mode as ``NA'' (no accretion, see Table 1).

Post-coalescence accretion is expected for gas-rich mergers, where MBH pairing and coalescence is believed to occur on very short timescales (Mayer et al. 2007). If we are to assume that during a galaxy merger the MBH remnant is always brought on the correct correlation with its host, by the combination of merging and accretion, two additional routes are possible.

\item Accretion is triggered {\it before} the MBHB coalescence and only the
more massive MBH ($M_1$) accretes mass; in this case the masses of the 
coalescing MBHs are $\alpha M_1$ and $M_2$, where 
\begin{equation}
\alpha=\frac{M_r-M_2}{M_1}-1\,.
\label{alpha}
\end{equation}   
We label this accretion mode as ``SA'' (single BH accretion, see Table 1).
\item Accretion is triggered {\it before} the MBHB coalescence and both MBHs are allowed to accrete the same fractional amount of 
mass; in this case the masses of the 
coalescing MBHs are $\beta M_1$ and $\beta M_2$, where 
\begin{equation}
\beta=\frac{M_r}{M_1+M_2}-1\,.
\label{beta}
\end{equation}   
We label this accretion mode as ``DA'' (double BH accretion, see Table 1).
\end{itemize}

\end{enumerate}
The "SA" and the "DA" modes are to be expected in gas-poor mergers, especially in non-equal-mass mergers, where the dynamical evolution of the binary is much slower (e.g., Yu 2002) than the infall timescale of the gas (e.g., Cox et al. 2008). In a stellar environment the orbital decay of MBHBs is expected to be much longer than in a gaseous environment (e.g., Sesana et al. 2007, Dotti et al. 2006). 

The MBHB models that we consider here relay on two assumptions: all bulges host a MBH, and a MBHB always coalesces following the hosts' merger. Regarding the first assumption, dynamical processes such as gravitational recoil and triple MBH interactions may deplete bulges from their central MBH. However, if one compares the mass function of coalescing binaries obtained here to the results of EPS merger tree models, where both gravitational recoil and triple interactions are consistently taken into account, one finds that the two distributions (shown in the upper--left panel of figure \ref{f1}) are in excellent agreement, within the statistical uncertainties. This is because triple interactions are likely to eject the lighter MBH from the host, leaving behind a massive binary in the vast majority of the cases (Volonteri Haardt \& Madau 2003, Hoffman \& Loeb 2007).  Gravitational recoil, on the other hand, may be effective in expelling light MBHs from protogalaxies at high redshift, but has probably a negligible impact on the population of MBHs in the mass range of interest for PTA observations (Volonteri 2007). The assumption that MBHBs coalesce within an Hubble time following the hosts' merger is justified by several recent studies of MBHB dynamics. The stellar distribution interacting with the binary may be efficiently repopulated as a consequence of non-axisymmetric or triaxial galaxy potentials (Merritt \& Poon 2004, Berzcik et al. 2006), or by massive perturbers (Perets \& Alexander 2008); moreover, if one considers post-Newtonian corrections to the binary evolution and the effects of eccentricity, one finds that the coalescence timescale is significantly reduced (Berentzen et al. 2008). If the binary evolution is gas--driven, typical hardening timescales are expected to be shorter than $10^8$ yr (Escala et al. 2005, Dotti et al. 2006). 

In summary, we build a total of 12 models (4 MBH-host prescriptions $\times$ 3 accretion modes). Hereinafter, we will refer to each model using the labels associated to the prescriptions employed in it. As an example, the model based on the $M_{\rm BH}-\sigma$ relation ("Tr") with single BH accretion prescription ("SA") will be referred to as Tr-SA, see Table 1 for a summary.

\subsection{Computing the coalescence distributions}

Assigning a MBH to each galaxy, we obtain a list of coalescences (labelled by MBH masses and redshift);
the same output quantity given by the EPS-based merger trees, used as a starting 
point in PaperI. Each event  in the list can be now properly weighted over the observable volume shell 
at each redshift to obtain the distribution $d^3N/d{\cal M}dzdt$ along the lines described in PaperI.

A further technical detail has to be considered to smooth the coalescence
distributions. The Millennium simulation provides better statistics for constructing
the mass function of merging objects, but the redshift sampling is rather poor. 
In fact, the Millennium database consists of 63 snapshots of the whole simulation. 
The most recent ones are taken at $z=0, 0.02, 0.04, 0.064$ and 0.089; 
we need, at least at low redshift, to spread 
the events over a continuum in $z$, to obtain a sensible 
distance distribution of the GW sources. To this aim, we decouple the 
$({\cal M},z)$ dependence in the differential mass function at $z<0.3$
and re-write ${d^3N}/{d{\cal M}dzdt}$ in the following form:
\begin{equation}
\frac{d^3N}{d{\cal M}dzdt}=\Psi(z)\times\int_{0.3}^0dz\frac{d^3N}{d{\cal M}dzdt}= \Psi(z)\times{\cal F}({\cal M},t)
\end{equation}   
By doing so we are redistributing according to a given function
$\Psi(z)$ the average mass function obtained at $z<0.3$. This is justified
since, as expected, the mass function does not show any significant 
evolution for redshifts below $0.3$. The dependence on $z$ should be of the form
$\Psi(z)\propto n_G^2\times dV_C/dz$, where $n_G$ is the galaxy/MBH number 
density and $dV_C/dz$ is the differential comoving volume shell. At such 
small redshifts the impact of merger activity on galaxy/MBH number density is 
negligible (of order of 0.1/Gyr, e.g., Wake et al. 2008; 
White et al. 2007; Masjedi et al. 2006; Bell et al. 2006); 
therefore, we assume $n_G$ to be constant. On the other hand the Universe can be 
considered Euclidean, so that the differential volume shell is just 
proportional to $z^2$. We then obtain the coalescing MBHB distribution in the form   
\begin{equation}
\frac{d^3N}{d{\cal M}dzdt}=C\times z^2\times{\cal F}({\cal M},t),
\label{coaldist}
\end{equation}   
where $C$ is a normalization factor set by the condition 
\begin{equation}
\int d{\cal M}\int_{0.3}^0 dz\frac{d^3N}{d{\cal M}dzdt}=\int d{\cal M}\int_{0.3}^0 dz\, C\times z^2\times{\cal F}({\cal M},t).
\end{equation}   

\begin{figure}
\centerline{\psfig{file=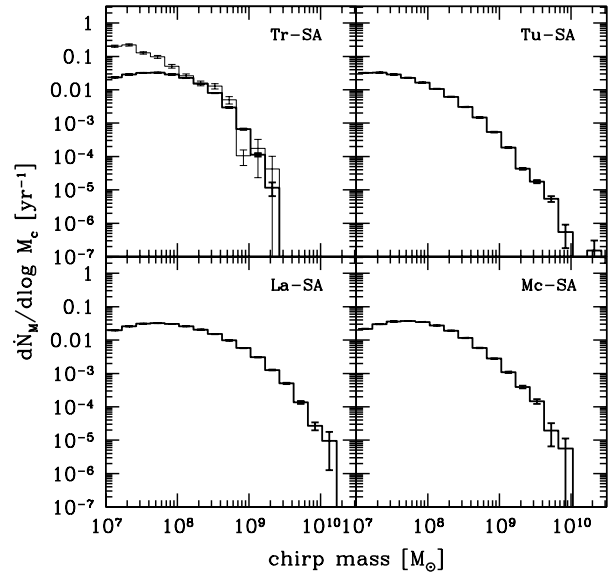,width=84.0mm}}
\caption{Mass function of coalescing MBHBs according to MBH--host
relations reported by several authors. The thick histograms refer to the model labeled
at the top of the panel and described in the text. Error bars are 
calculated assuming a Poisson error in the number count of events 
from the coalescence catalogues that contribute to the chirp mass 
interval. In the top-left panel the thin histogram
is the coalescing MBHB mass function predicted by the VHMhopk
model studied in PaperI.}
\label{f1}
\end{figure}

The ${\cal M}$ distributions of coalescing binaries are shown in figure \ref{f1}
for all the MBH-host prescriptions assuming accretion on $M_1$ only (models Tr-SA, Tu-SA, Mc-SA, La-SA ). 
The top-left panel also shows the distribution obtained by a reference
EPS merger tree model  (VHMhopk; Volonteri, Salvaterra \& Haardt 2006) used in PaperI. The agreement with the $M_{\rm BH}-\sigma$
prescription ("Tr") is good for ${\cal M} > 10^8\Ms$, although the statistical errors are large due to low number statistics. 
The discrepancy for ${\cal M}<10^8\Ms$
is due both to the resolution limit of the Millennium simulation and to the fact
that we relate MBHs to bulges. However, the fact that our sample may be incomplete for 
${\cal M}<10^8\Ms$ has little (if any) impact on the results of this paper: the MBHB population
observable with PTAs is by far dominated by sources with ${\cal M}>10^8\Ms$, as we have shown
in PaperI, and even more so when we consider systems that can be individually resolved. We will
further discuss this point in Section 4. 
Note that, as discussed by, e.g., Lauer et al. 2007 and Tundo et al. 2007, the high mass end of the population derived using the $M_{\rm BH}-\sigma$ relation ("Tr" models) drops very steeply. The drop is much faster than in all the other cases, that is for distributions inferred from the $M_{\rm BH}-M_{\rm bulge}$ or the $M_{\rm BH}-M_V$ relations ("Tu", "Mc", "La" models). This is because $\sigma$ seems to converge to a plateau in the limit of very massive galaxies (Lauer et al. 2007).

\section{Timing residuals from resolved massive black hole binaries}

The search for GWs using timing data exploits the effect of gravitational radiation on the propagation of the radio waves from one (or more) pulsar(s). The characteristic signature of GWs on the time of arrival (TOA) of radio pulses (e.g. Sazhin 1978, Detweiler 1979, Bertotti et al. 1983) is a linear combination of the two independent GW polarisations. In practice, the analysis consists in computing the difference between the expected and actual TOA of pulses; the {\em timing residuals} contain information on all the effects that have not been included in the fit, including GWs. In this section we summarise the observed signal produced by GWs, following closely Jenet et al. (2004), to which we refer the reader for further details.

The observed timing residual generated by a GW source described by the independent polarisation amplitudes $h_{+,\times}$ is
\be
r(t) = \frac{1}{2} (1 + \cos\mu) \left[r_+(t) \cos(2\psi) + r_\times(t) \sin(2\psi)\,,
\right]
\label{e:R}
\ee
where $t$ is the time at the receiver, $\mu$ is the opening angle between the GW source and the pulsar relative to Earth and $\psi$ is the GW polarisation angle. The two functions $r_{+,\times}(t)$ are defined as
\ba
r_{+,\times}(t) & = & r^{(e)}_{+,\times}(t) - r^{(p)}_{+,\times}(t)\,,
\label{e:r}
\\
r^{(e)}_{+,\times}(t) & = & \int_0^t dt' h^{(e)}_{+,\times}(t')\,,
\label{e:re}
\\
r^{(p)}_{+,\times}(t) & = & \int_0^t dt' h^{(p)}_{+,\times}\left[t' - \frac{d}{c} (1 - \cos\mu)\right]\,.
\label{e:rp}
\ea
Note that $r^{(e)}_{+,\times}(t)$ and $r^{(p)}_{+,\times}(t)$ have the same functional form, and result from the integration of the time evolution of the polarisation amplitudes at different times, with a delay $\simeq 3.3 \times 10^3\,(d/1\,\mathrm{kpc}) (1 - \cos\mu)$ yr, where $d$ is the distance of the pulsar from the Earth. For GWs propagating exactly along the Earth-pulsar direction, there is no effect on the TOAs ($r(t) = 0$ for $\cos\mu = \pm 1$).

From now on we will concentrate on the timing residuals produced by binary systems in {\em circular} orbit. We model gravitational radiation at the leading quadrupole Newtonian order, that is fully justified by the fact that binaries in the mass and frequency range considered here are far from the final merger; in fact, the time to coalescence is $\simeq 615 ({\cal M}/10^9\,\Ms)^{-5/3}\,(f/5\times 10^{-8}\,\mathrm{Hz})^{-8/3}$ yr. The timing residuals~(\ref{e:R}) can be written as (Jenet et al, 2004):
\be
r(t) = r^{(e)}(t) - r^{(p)}(t)
\label{e:R1}
\ee
where
\ba
r^{(e)}(t) = \alpha(t) \left[a_+ (1+ \cos^2\iota) \cos\Phi(t) + 2 a_\times \cos\iota \sin\Phi(t)\right]
\label{e:Re}
\ea
In the previous expressions $\iota$ is the source inclination angle, $\Phi(t)$ the GW phase related to the frequency $f(t)$ (twice the orbital frequency) by 
\be
\Phi(t) = 2 \pi \int^t f(t')dt'\,,
\label{e:gwphase}
\ee
and
\ba
\alpha[f(t)] & = & \frac{{\cal M}^{5/3}}{D}\,\left[\pi f(t)\right]^{-1/3}
\nonumber\\
& \simeq & 25.7\, \left(\frac{{\cal M}}{10^9\,\Ms}\right)^{5/3}\,\left(\frac{D}{100\,\mathrm{Mpc}}\right)^{-1}
\nonumber\\
&&\times \left(\frac{f}{5\times 10^{-8}\,\mathrm{Hz}}\right)^{-1/3}\,\mathrm{ns}
\label{e:alpha}
\ea
is an overall scale factor that sets the size of the residuals; $D$ is the luminosity distance to the GW source. The expression for $r^{(p)}(t)$ can be simply obtained from Equations~(\ref{e:Re}), (\ref{e:gwphase}) and~(\ref{e:alpha}) by shifting the time $t \rightarrow t - d (1 - \cos\mu)/c$, see equation (\ref{e:re}) and~(\ref{e:rp}). The two functions $a_+$ and $a_\times$ are the "antenna beam patterns" that depend on the source location in the sky and the polarisation of the wave.

MBHBs observable with PTAs produce a quasi-monochromatic signal -- the frequency change is $\simeq 3\times 10^{-2} ({\cal M}/10^9\,\Ms)^{5/3}\,(f/5\times 10^{-8}\,\mathrm{Hz})^{11/3}$ nHz $\mathrm{yr}^{-1}$ -- of known form (though unknown parameters). The optimal data analysis approach to search for these signals is by means of the well known technique of matched-filtering. The data set can be represented as:
\be
\delta t(t) = r(t) + \delta t_n(t)
\ee
where $r(t)$ is the contribution from GWs, and $\delta t_n(t)$ accounts for the fluctuations due to noise; the latter contribution is the superposition of the intrinsic noise in the measurements and the GW stochastic background from the whole population of MBHBs. The angle-averaged optimal signal-to-noise ratio at which a signal from a MBHB radiating at (GW) frequency $\approx f$ can be detected using a {\em single} pulsar is
\be
\langle \rho^2 \rangle= \left[\frac{\delta t_\mathrm{gw}(f)}{\delta t_\mathrm{rms}(f)}\right]^2\,.
\label{e:snr}
\ee
In the previous expression $\delta t_\mathrm{rms}(f)$ is the root-mean-square value of the noise level $\delta t_n$ at frequency $f$, $\langle .\rangle$ represents the average over the source position in the sky and orientation of the orbital plane, and $\delta t_\mathrm{gw}(f)$ is the characteristic amplitude of the timing residual over the observation time $T$ defined as:
\be
\delta t_\mathrm{gw}(f) = \frac{8}{15} \alpha(f) \sqrt{fT}\,,
\label{e:deltatgw}
\ee
where the numerical pre-factor comes from the angle average of the amplitude of the signal:
\be
\left\langle a_+^2 (1+ \cos^2\iota)^2 + 4 a_\times^2 \cos^2\iota \right\rangle = \frac{8}{15}\,.
\label{e:geometryaverage}
\ee

Equation~\ref{e:snr} is appropriate to describe observations using a single pulsar. In reality one can take advantage of the several pulsars that are continuously monitored to increase the signal-to-noise ratio, and therefore the confidence of detection: adding coherently the residuals from several pulsars -- currently the Parkes PTA contains about 20 pulsars, and more are expected to available with future instruments -- yields an increase in signal-to-noise ratio proportional to the square-root of the number of pulsars used in the observations. We will use the characteristic amplitude of the residuals $\delta t_\mathrm{gw}$ to quantify the strength of a GW signal in PTA observations; $\delta t_\mathrm{gw}$ can be used to compute in a straightforward way the signal-to-noise ratio, as a function of the noise level and number of pulsars in the array (all of which are quantities that do not depend on the astrophysical model), and therefore asses the probability of detection of sources in the context of a given MBHB assembly scenario.

\section{Results}

For each of the twelve models considered in Section 2 -- the models are the result of four MBH-host galaxy prescriptions and three different accretion scenarios --  that encompass a very broad range of
MBHB's assembly scenarios, we compute the number of sources that are potentially resolvable individually and several statistical properties of the population, such as the redshift, chip-mass and frequency distributions, by means of Monte-Carlo realisations of the whole population of MBHBs according to a given model. Before describing the results, we provide details about each step in the computation of the relevant quantities. 

The distribution given by equation (\ref{coaldist}) is straightforwardly converted into 
$d^3N/dzd{\cal M} d{\rm ln}f_r$, i.e. the comoving number of binaries 
emitting in a given logarithmic (rest-frame) frequency interval with chirp mass and 
redshift in the range $[{\cal M},{\cal M}+d{\cal M}]$ and $[z, z+dz]$, 
along the lines described in Section 3 of PaperI. 
As in PaperI, we then assume that in the frequency range of interest ($f>3\times 10^{-9}$ Hz) the binary evolution is driven by GW emission only. This is a reasonable assumption, since the coalescence timescale for these systems is typically $\lsim10^6$ yr; any other putative mechanism (i.e. star ejection, gas torques) of angular momentum removal must have an enormous efficiency to compete with radiation reaction on such short timescales. 
For each of the twelve models we estimate the GW stochastic background 
(and the corresponding rms value of the timing residuals as a function of frequency) generated 
by the sources following the scheme described in Section 4 of PaperI.
Finally we generate distribution of bright, individually resolvable sources 
by running 1000 Monte-Carlo realisations of the whole population of MBHBs and by 
selecting only those sources whose characteristic timing residuals, equation (~\ref{e:deltatgw}), 
exceed the stochastic background level.
We note that the result of which and how many sources raise above the average noise level depends on the duration of the observation $T$, for two reasons: (i) $T$ affects the size of the observational window in frequency space, in particular the minimum frequency $1/T$ that can be reached, and (ii) the background level decreases as the observation time increases (as the size of the frequency resolution bin $\Delta f=1/T$ decreases), enhancing the number of individually resolvable sources. For the results that are described here and summarised in table \ref{tab:summary}, we set $T = 5$ yrs; increasing the data span to 10 years, the background level would be slightly lower, and the additional resolved sources would be barely brighter than the background. The statistics of bright, well resolvable sources is basically unaffected. We can then cast the results in terms of the cumulative number of resolvable sources as a function of the timing residuals, according to:
\be
N(\delta t_\mathrm{gw}) = \int_{\delta t_\mathrm{gw}}^\infty \frac{d N}{\delta t_\mathrm{gw}'}\delta t_\mathrm{gw}'\,,
\label{e:N}
\ee
where the integral is restricted to the sources that produce residuals above the rms level of the stochastic background; if we do not consider this additional constraint, then, for any given $\delta t_\mathrm{gw}$ , $N(\delta t_\mathrm{gw})$  simply returns the total number of sources in the Monte-Carlo realisation above that particular residual threshold. 

Each Monte-Carlo realisation clearly yields a different value for $N(\delta t_\mathrm{gw})$ (or its distribution according to a given parameter); the values quoted in the next section and summarised in table \ref{tab:summary} refer to the sample mean computed over the set of Monte-Carlo realisations and the sample standard deviation. Figure \ref{f5b} quantifies the typical 1-$\sigma$ error in our estimate of the number of sources.

\subsection{Single resolvable binaries}

\begin{figure}
\centerline{\psfig{file=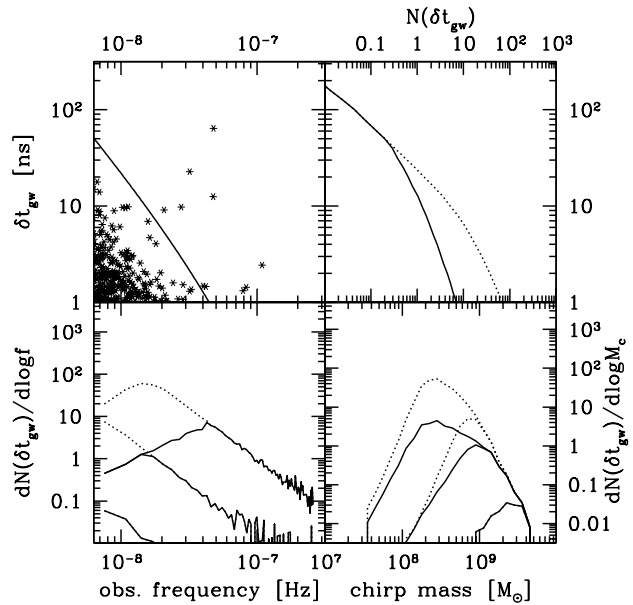,width=84.0mm}}
\caption{Summary of the properties of the population of massive black hole binary systems -- according to model Tu-SA -- that generate gravitational waves in the frequency window covered by Pulsar Timing Arrays. {\it Top-left panel}: characteristic amplitude of the 
timing residuals $\delta t_\mathrm{gw}$ (equation(~\ref{e:deltatgw})) as a function of frequency; the asterisks are the residuals generated by individual sources and the solid line is the estimated level of the GW stochastic background. {\it Top-right panel}: 
$\delta t_\mathrm{gw}$ as a function of the number $N(\delta t_\mathrm{gw})$ of total (dotted line) and individually resolvable (solid line) sources, see equation \ref{e:N}.  {\it Bottom-left panel}: Distribution of the number of total (dotted lines) and resolvable (solid lines) sources per logarithmic frequency interval $dN(\delta t_\mathrm{gw})/d\log f$ as a function of the GW frequency for different values of $\delta t_\mathrm{gw}$: from top to bottom 1, 10 and 100 ns, respectively. {\it Bottom-right panel}: distribution of the total (dotted lines) and individually resolvable (solid lines) number of sources per logarithmic chirp mass interval $dN(\delta t_\mathrm{gw})/d\log {\cal M}$ as a function of chirp mass for different values of $\delta t_\mathrm{gw}$: from top to bottom 1, 10 and 100 ns, respectively. An observation time of 5 years is assumed.}
\label{f4}
\end{figure}

   \begin{figure*}
   \centering
   \resizebox{\hsize}{!}{\includegraphics[clip=true]{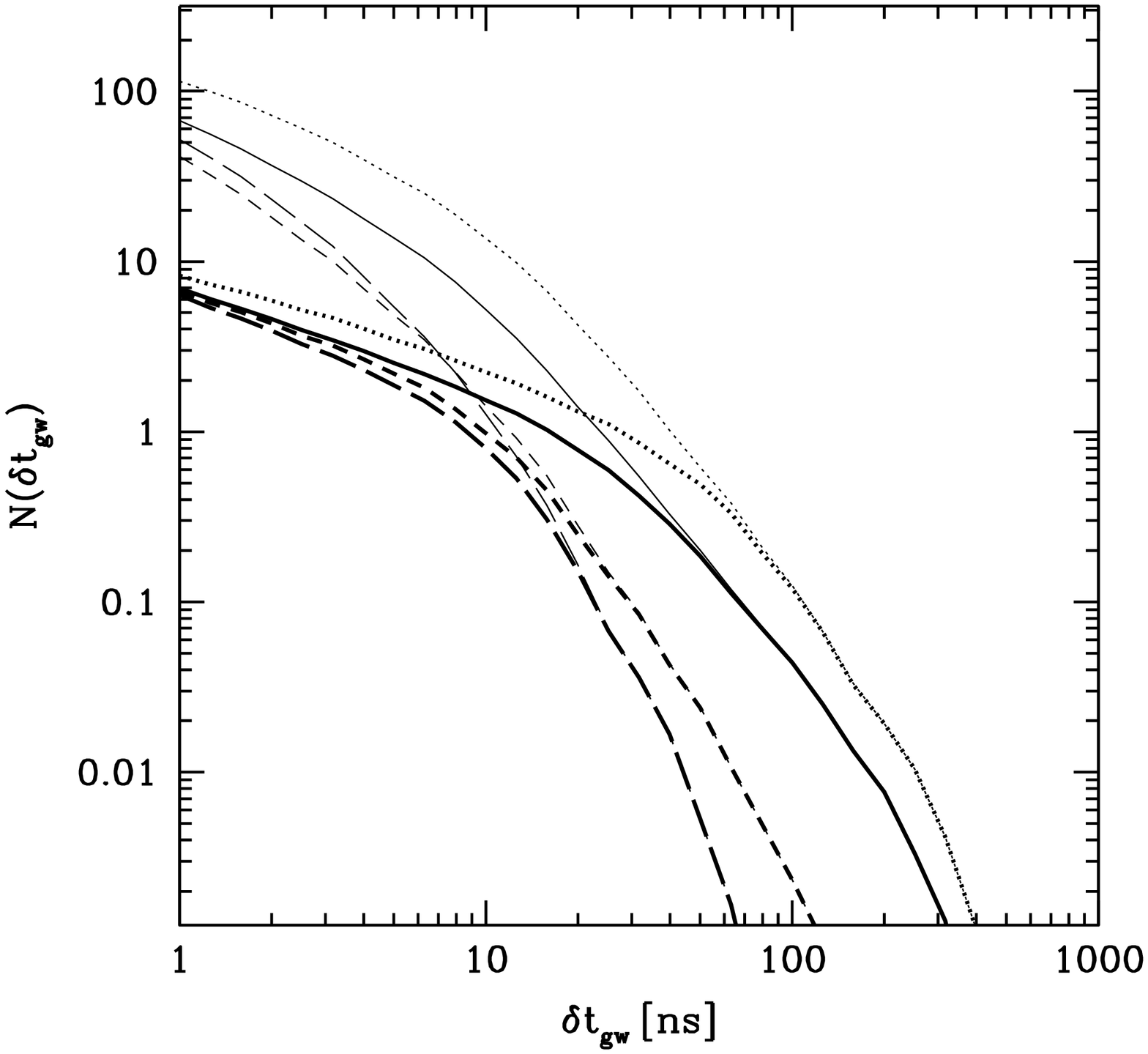}
   \includegraphics[clip=true]{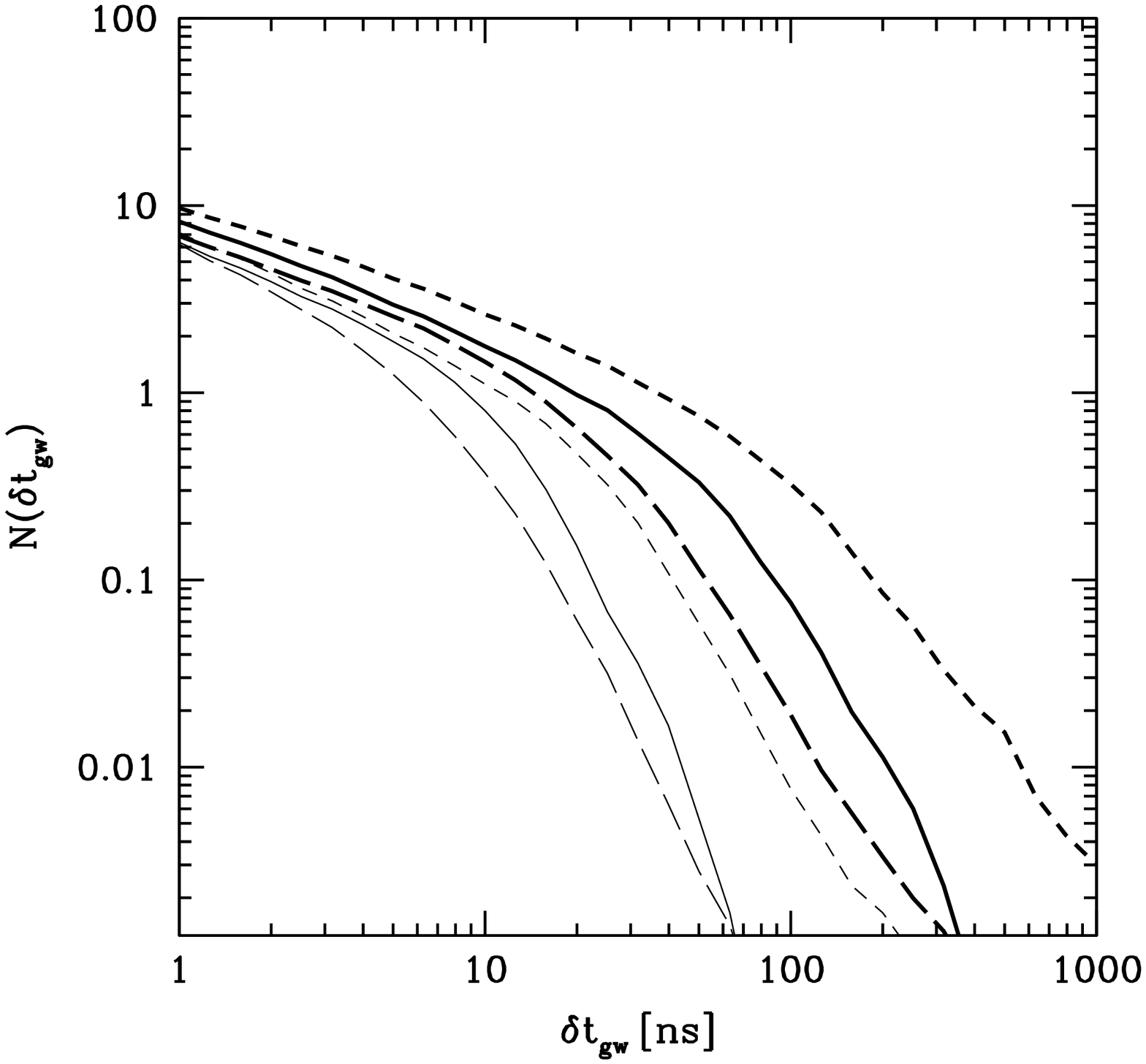}}
     \caption{{\it Left panel}: The effect of the MBH-host galaxy relation, assuming that accretion always takes
place for a single black hole before merger ("SA" models), 
on the number of observable systems. The plot shows 
the number of total (thin lines) and resolvable
(thick lines) sources $N(\delta t_\mathrm{gw})$ as a function of 
$\delta t_\mathrm{gw}$, see equation \ref{e:N}. Four different MBH merger scenarios are
considered: Tu-SA (solid line), Tr-SA (long--dashed lines), Mc-SA (short--dashed lines) and La-SA (dotted lines), 
see Section 2 for a description of the models. {\it Right panel}: 
The effect of the MBH accretion model on the number of individually observable systems. 
The plot shows the number of resolvable sources only, $N(\delta t_\mathrm{gw})$ as a function of 
$\delta t_\mathrm{gw}$. As reference for the MBH-host galaxy relation, models "La" (thick lines) and "Tr" (thin lines) are considered. The line style is as follow: model La-SA and Tr-SA (solid lines), La-DA and Tr-DA (short--dashed lines), and 
La-NA and Tr-NA (long--dashed lines). The duration of the observation is set to $T = 5$ yr}  
        \label{f5}
    \end{figure*}

The large number of Monte-Carlo realisations allows us to study the details of the properties of the
individual sources in a statistical sense. We concentrate in particular on the physical properties
of the population, such as the expected number of sources per logarithmic frequency interval 
$dN(\delta t_\mathrm{gw})/d\log f$, chirp mass range $dN(\delta t_\mathrm{gw})/d\log {\cal M}$
and redshift $dN(\delta t_\mathrm{gw})/dz$, and the 
observable parameters, such as the timing residuals produced by each system and
the overall expected number of resolvable MBHBs  at a given level of timing residual noise. 
A summary of the typical range of information that can be extracted from the simulations is
shown in figure \ref{f4} for the specific model Tu-SA. The top-left panel shows the induced 
residuals of each single source compared to the level produced by the stochastic background
from the whole population; the plot clearly shows the importance of taking
into account the additional "noise contribution" from the brightness of the GW sky in considering
the detectability of resolvable systems. There are many sources inducing residuals above, say the
5 ns level, however most of them contribute to the build-up of the background, 
and are not individually resolvable. 
The expected number of bright resolvable MBHBs at 
frequencies $<10^{-7}$Hz, and at a timing level $>1$ns, is typically around ten, with fainter 
sources resolvable at higher frequencies. The top-right panel shows
the mean number of individual sources detectable as a function of $\delta t_\mathrm{gw}$ from
1000 Monte-Carlo realizations of the emitting population. In this particular case
a sensitivity of $\approx 10$ ns is required to resolve an individual source; for
a timing precision of 100 ns there is a 5\% chances to observe a particularly bright
source. Note that at the 1 ns level, there are $\sim 100$
MBHBs contributing to the signal, however 90\% of them contribute to the background and
only about 10 sources are individually resolvable. In the two bottom panels of
figure \ref{f4} we plot the frequency and chirp mass distributions of resolvable sources
for selected values of the minimum detectable residual amplitude $\delta t_{\rm gw}$. 
Not surprisingly, the chirp mass of observable systems decreases for smaller values of
the considered residual threshold, however even for an rms level of 1 ns all the systems are characterised by 
${\cal M}>10^8\msun$.
The frequency distribution shows instead a peak corresponding to the frequency at which
the background level equals the selected value of $\delta t_{\rm gw}$. At higher frequencies
the number of sources drastically drops because the number of emitting binaries 
is a quite steep function of frequency, $N(f)\propto f^{-8/3}$; the decrease at lower
frequencies is because most of the emitters actually contribute to the background and
are not individually resolved (as clearly shown by the dotted lines). The qualitative behaviour of
the results obtained using different astrophysical models is very similar to the one 
described in figure~\ref{f4}, with differences that affect only the numerical values of the
different quantities.

\begin{figure}
\centerline{\psfig{file=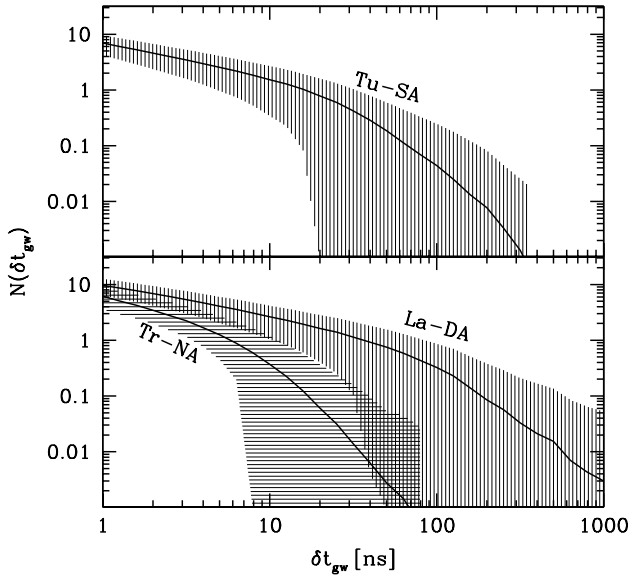,width=84.0mm}}
\caption{The number of individually resolvable sources for selected MBHBs assembly models. The plots
show $N(\delta t_\mathrm{gw})$ as a function of $\delta t_\mathrm{gw}$, see equations ~\ref{e:N} and~\ref{e:deltatgw}, for a typical 
model (Tu-SA, top panel), and, in the lower panel, the two models that yield the largest (La-DA: top curve) and 
smallest (Tr-NA: bottom curve) number of sources. The solid lines represent
the mean value of $N(\delta t_\mathrm{gw})$ and the shaded area the $1-\sigma$ region as computed
from 1000 Monte-Carlo realisations of each MBHB population, see also Table 1.}
\label{f5b}
\end{figure}
\begin{figure}
\centerline{\psfig{file=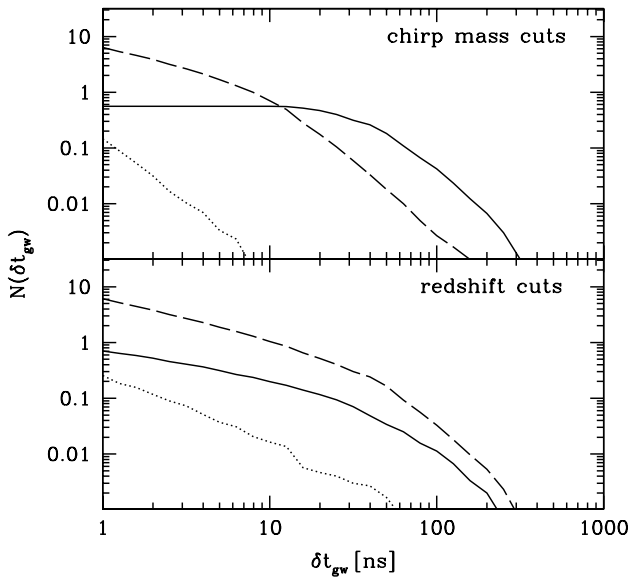,width=84.0mm}}
\caption{{\it Top panel} Cumulative mean number of resolvable
sources $N(\delta t_\mathrm{gw})$ as a function of the characteristic timing residual $\delta t_\mathrm{gw}$ for different mass cuts:
solid line: ${\cal M}>10^9\msun$; dashed line: $10^8\msun<{\cal M}<10^9\msun$;
dotted line $10^7\msun<{\cal M}<10^8\msun$.{\it Bottom panel}: same 
as top panel but for different redshift intervals: solid line $z<0.1$; dashed
line $0.1<z<1$; dotted line $z>1$.}
\label{f6}
\end{figure}
\begin{figure}
\centerline{\psfig{file=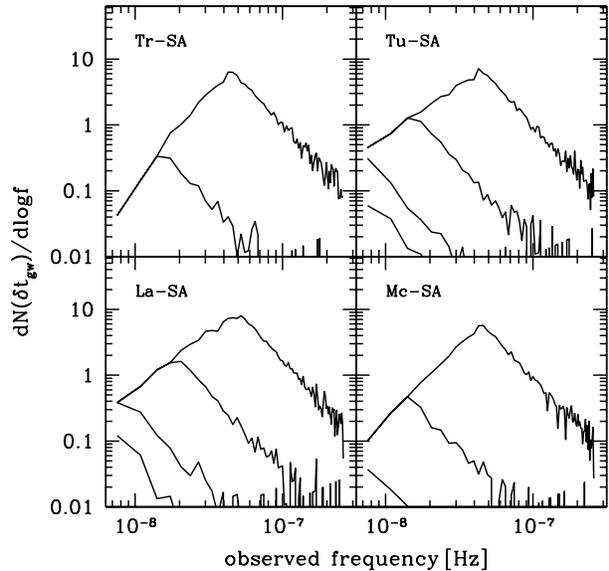,width=84.0mm}}
\caption{Distribution of the number of resolvable sources as a function of
the gravitational wave frequency. 
The plots show $dN(\delta t_\mathrm{gw})/d\log f$ for different values of
the characteristic amplitude of the timing residuals 
(from right to left $\delta t_\mathrm{gw}=1, 10, 50$ and 100 ns).
Each panel refers to
a different  MBH-host relation, while SA mode is considered 
(see labels in each panel). Chirp mass and
redshift distributions for the same models are given in 
figures~\ref{f8} and~\ref{f9}.}
\label{f7}
\end{figure}
\begin{figure}
\centerline{\psfig{file=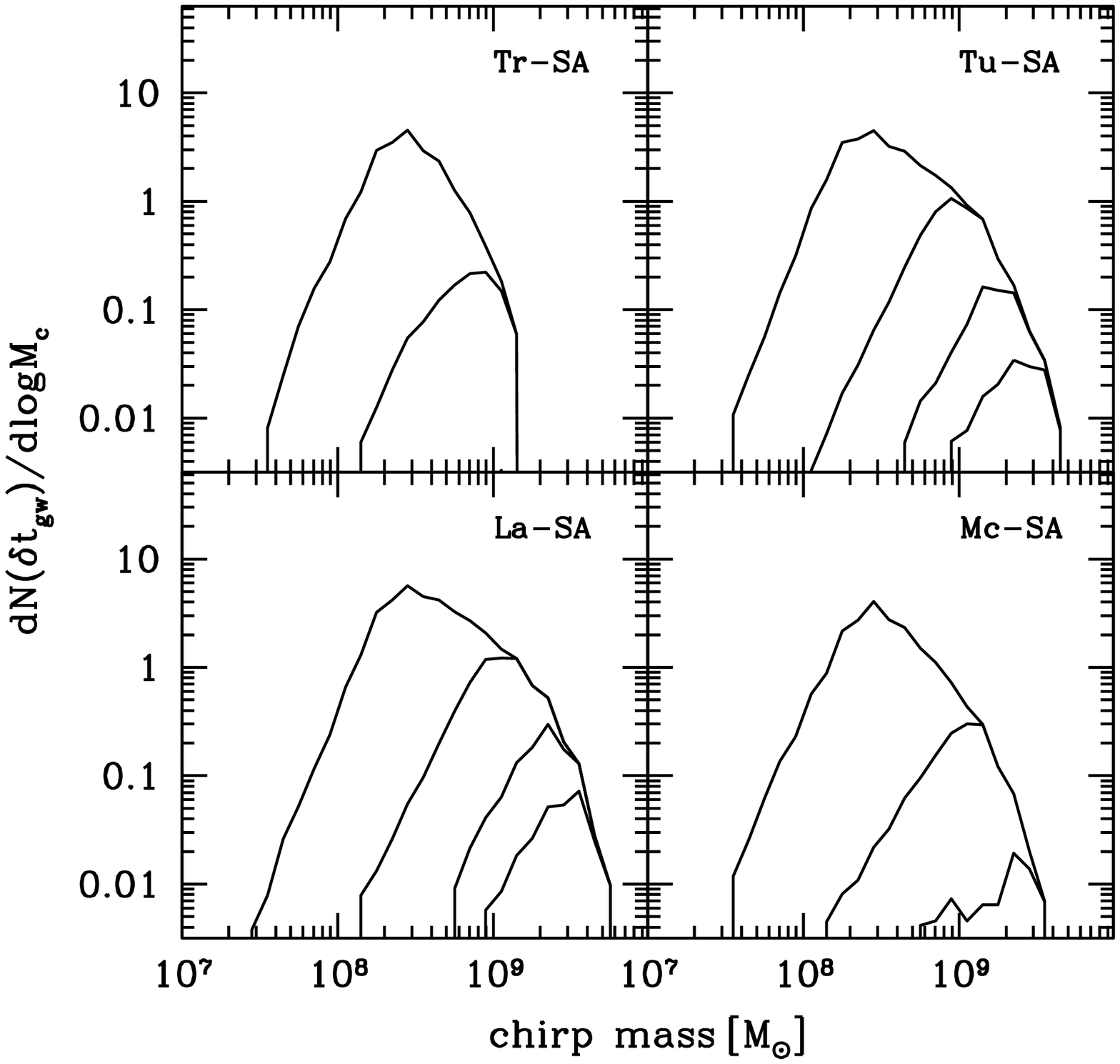,width=84.0mm}}
\caption{Chirp mass distribution of the number of resolvable sources. 
The plots show $dN(\delta t_\mathrm{gw})/d\log {\cal M}$ 
for the same characteristic amplitude of the timing residuals and 
models as in figures~\ref{f7} and ~\ref{f9}.
}
\label{f8}
\end{figure}
\begin{figure}
\centerline{\psfig{file=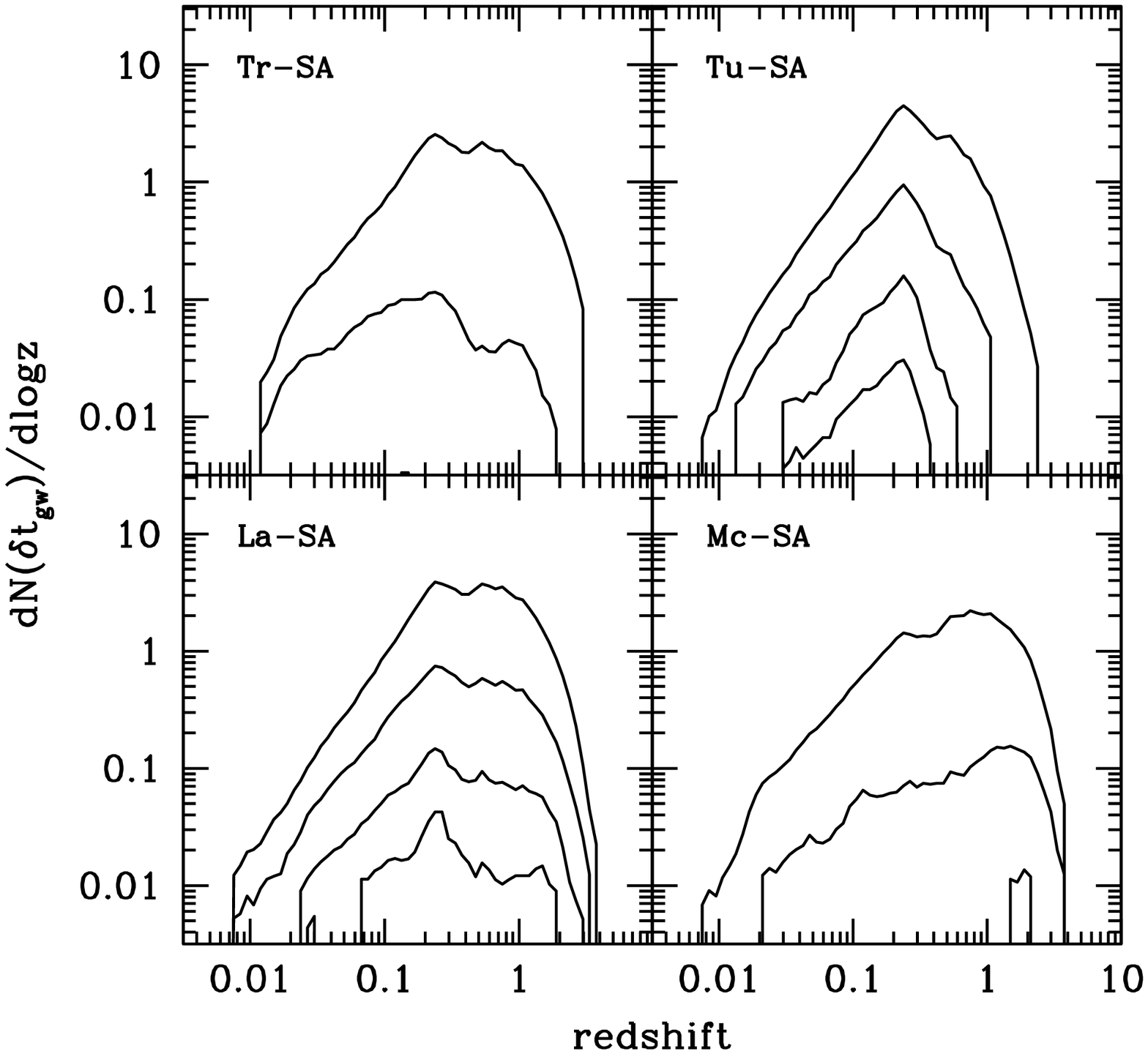,width=84.0mm}}
\caption{Redshift distribution of the resolvable sources. 
The plots show $dN(\delta t_\mathrm{gw})/d\log z$ 
for the same characteristic amplitude of the timing residuals and 
models as in figures~\ref{f7} and~\ref{f8}.}
\label{f9}
\end{figure}

The central question that we want to address in this paper is what is the 
expected number of individually resolvable sources that produce an effective
timing residual above a given value, as a function of different models of MBHB
formation and evolution. We summarise these results in figures \ref{f5} and \ref{f5b},
and in Table~\ref{tab:summary},
where we show the mean total number of individual sources that exceed a given 
level of timing residual (as defined by equation (\ref{e:N})), as a function of the timing residual. 
The qualitative behaviour
of the results is similar for all the scenarios, but the actual numbers
vary significantly. In fact, both MBH-host relations and accretion prescriptions have a strong 
impact on the statistics of the bright sources and, consequently, on their detection.
We analyse the effect of the black holes populations and of the accretion in turn. The left panel of figure
\ref{f5} shows results where the accretion prescription is the same ("SA"),
but the underlying MBH-host relation changes; on the other hand, in the right panel, 
we select two MBH-host relations, but change the accretion history. The
left panel shows that
assuming a sensitivity threshold of 30 ns, one expects to observe of the order of one
source in the La-SA model, while there is only a probability $\approx$ 5\% for the Tr-SA model. 
If we maintain the same MBH-host relation and we consider 
different accretion scenarios for {\em e.g.} the Lauer et al. population, the mean number of expected 
sources varies by a factor of $\approx 5$ between $0.3$ (La-NA) and $1.5$ (La-DA), see right panel. Figure \ref{f5}
shows that in the most pessimistic case -- Tr-NA model, that has the sharpest cut--off of 
the MBHB mass function bright end -- a precision of $\approx 5$ns should guarantee a positive
detection; in the optimistic La-DA case, a precision of $\approx 50$ns should be sufficient.
In turn the timing precision required for positive detection is in the range $5-50$ns,
that is basically consistent with a factor $\approx 10$ uncertainty in the background
level estimated in PaperI.

The typical spread around the mean values obtained in the Monte-Calro realisations is shown in figure~\ref{f5b} for selected models. When $N(\delta t_\mathrm{gw})\gg 1$ the 1-$\sigma$ range is roughly the Poisson error around 
the mean (reflecting the uncorrelated nature of sources in each Monte Carlo realisation).  
When $N(\delta t_\mathrm{gw})<1$ it can be interpreted as the probability to find a single
source above the considered $\delta t_\mathrm{gw}$ value if the actual MBHB population in the Universe follows the prescription given by the considered model; in this case a non-detection is 
trivially consistent with the model predictions. Table 1 provides a summary of the results for the 12 models.

It is also interesting to investigate the mass-redshift distribution of the detectable sources. 
Figure \ref{f6} shows the expected number of detectable individual sources (as a function 
of the residuals amplitude) for different redshift and mass ranges. Obviously,
higher timing residuals correspond to higher ${\cal M}$, since the strength of the signal is
proportional to ${\cal M}^{5/3}$, and the most likely sources to be detected fall in the
range $10^8 \msun \simlt {\cal M} \simlt 10^9 \msun$ (MBHBs with ${\cal M} > 10^9 \msun$
produce indeed larger timing residuals, but are also much rarer). Interestingly, the 
vast majority of detectable sources are at redshift $0.1 \simlt z \simlt 1$, which shows
that PTAs could probe the medium-redshift Universe, and are unlikely to discover
nearby sources. The reason is simply that, at least at small redshift, the Universe volume
increases as $z^3$. 
 
A summary of the properties of individual resolvable sources, $dN(\delta t_\mathrm{gw})/d\log f$, 
$dN(\delta t_\mathrm{gw})/d\log {\cal M}$ and $dN(\delta t_\mathrm{gw})/d\log z$,  is given in 
figures \ref{f7}, \ref{f8} and \ref{f9}, respectively, for all the four MBHB population models considered
here, with accretion limited to a single black hole prior to merger 
considering different residual thresholds $\delta t_{\rm gw}=1, 10, 50, 100$ ns.
All the models show the same qualitative features, as we have highlighted before. 
The frequency distribution shown in figure \ref{f7} was discussed above and
is the same for all the models. The distribution of the detectable sources as a function
of chirp mass (figure \ref{f8}) peaks at $\approx 3\times 10^8\msun$ for all models assuming 
$\delta t_{\rm gw}=1$ ns. Increasing $\delta t_{\rm gw}$, the distribution peak shifts towards higher   
${\cal M}$ and the mean number of events is strongly model dependent for 
$\delta t_{\rm gw}>10$ ns, cf. the values in Table \ref{tab:summary}. 
The redshift distributions (figure \ref{f9}) consistently show
a broad peak in the range $0.2<z<1$, due to the volume effect previously discussed.
Note that in the ``Mc" model the peak shifts toward higher redshifts as  
$\delta t_{\rm gw}$ increases, because in this model similar galaxies are 
populated by more massive black holes if found at higher redshifts, see
equation (\ref{mclure}).

\subsection{Stochastic background}

\begin{figure}
\centerline{\psfig{file=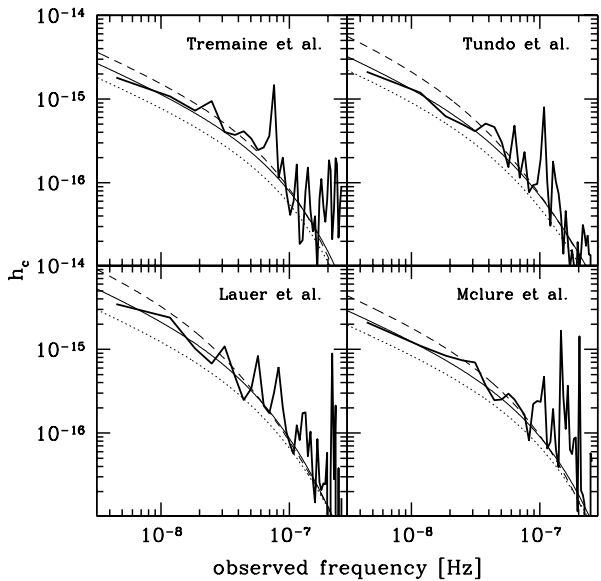,width=84.0mm}}
\caption{The characteristic amplitude of the 
GW stochastic background from the population of
MBHB systems. In
each panel the thin lines identify the estimated level of the stochastic background assuming
``SA" (solid line), ``DA'' (dashed line) and ``NA'' (dotted line) accretion modes. 
The total GW amplitude from a single Monte--Carlo realisation
of the signal corresponding to the ``SA'' accretion mode is also 
shown as thick solid line.}
\label{f2}
\end{figure}
\begin{figure}
\centerline{\psfig{file=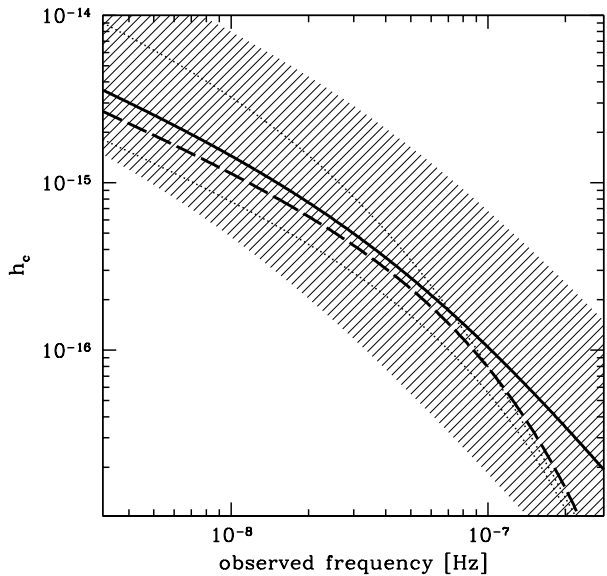,width=84.0mm}}
\caption{The characteristic amplitude of the GW stochastic background
compared to the estimate given in PaperI. The thick dashed line is the stochastic background
predicted by the Tr-SA model; the dotted lines bound the background levels computed for 
all the models investigated in this paper. For comparison, the solid thick line is the background predicted by the VHMhopk model and the shaded area the range of uncertainty of the strength of the
signal, as reported in PaperI (see Sections 4 and 5 of PaperI for details).}
\label{f3}
\end{figure}

As a sanity check, we compare the stochastic backgrounds derived according to the
MBH populations inferred using the Millennium Simulation to the predictions of 
EPS--based models reported in PaperI (the reader is referred to PaperI for the technical 
details). In figure \ref{f2} we show Monte Carlo generated
signals for each model; in each panel we plot the stochastic levels according
to the three different accretion modes discussed in Section 2. Both the accretion 
prescription and the adopted MBH-host correlation influence the level of the 
background. If accretion occurs onto the remnant (i.e. after coalescence, "NA" models), the 
predicted characteristic amplitude of the GW background can be up to a factor of 
3 lower with respect to models in which both the MBHs accrete before the final coalescence ("DA" models); 
on the other hand, $M_{\rm BH}-\sigma$ ("Tr") models predict lower backgrounds compared to
$M_{\rm BH}-M_{\rm bulge}$ ("Tu", "Mc") and $M_{\rm BH}-M_V$ ("La") models. 
A comparison of these results with those presented in PaperI is given in figure \ref{f3}. 
At $10^{-8}$ Hz the models studied here cover a characteristic amplitude range consistent with the 
uncertainty estimated in PaperI. The Tr-SA model predicts
a stochastic background that agrees with the typical EPS--model within
30\% for $f<10^{-7}$Hz. The high frequency end is instead steeper; 
this effect is caused by incompleteness in the low mass end of the MBH population.
As shown in figure \ref{f1}, the mass function of coalescing MBHB derived from the Millennium
simulation is not consistent with the one obtained using the EPS formalism  
for ${\cal M}<10^8\msun$. The weight of a single dark matter
particle in the Millennium simulation is $8.66/h_{100}\times10^8\msun$, allowing
the reconstruction of haloes with minimum mass of the order of 
$\approx5\times10^{10}\msun$. Assuming a barionic fraction of 0.1, the simulation
is then incomplete for barionic structures less massive than 
$\approx5\times10^{9}\msun$. We checked this by plotting the mass function
of barionic structures and finding a sudden drop below $10^9\msun$. It is then
inevitable that in the results derived from the Millennium Simulation most of the 
MBHs with mass below a few$\times10^6\msun$ are missing. 
Since many of these MBHs are expected to merge with more massive
ones during cosmic history, the (spurious) lack of MBHs in this mass rage
explains the flattening of the mass function $d\dot{N}_M/d\log {\cal M}$
shown in the top-left panel of figure \ref{f1}.
All the backgrounds are rather similar at $f>10^{-7}$ Hz because
all the MBH prescriptions adopted lead to similar MBH mass functions at 
$M_{\rm BH}<10^8\msun$ (this fact is independent of the incompleteness
issue). This means that the slope of the background 
on the right of the knee has a well defined dependence upon the adopted
MBH population: the more pronounced is the high mass tail of the MBH
mass function, the steeper is the high frequency end of the GW background.
In turn, models constructed using the Millennium simulation confirm 
the findings of PaperI.

\section{Summary and Conclusions}

We have investigated the ability of Pulsar Timing Arrays (PTAs) to resolve individual massive black hole binary systems by detecting gravitational radiation produced during the in-spiral phase through its effect on the residuals of the time of arrival of signals from radio pulsars. We have considered a broad range of assembly scenarios, using the data of the Millennium simulation to evaluate the galactic haloes merger rates, and a total of twelve different models that control the relations between the mass of the central black holes and the galactic haloes, and the evolution of the black hole masses through accretion. These models therefore cover qualitatively (and to large extent quantitatively) the whole spectrum of MBH assembly scenarios currently considered. Regardless of the model, we estimate that at least one resolvable source is expected to produce timing residuals in the range $\sim5-50$ ns, and therefore future PTAs, and in particular the Square Kilometre Array may be able to observe these systems. A whistle-stop summary of the models and results is contained in Table~\ref{tab:summary}. The total number of visible events clearly depend on the sensitivity of PTAs and on the astrophysical scenario. As expected, the brightest sources (for PTAs) are very massive binaries with chirp mass ${\cal M}>5\times10^8\msun$. However, (initially) quite surprisingly most of the resolvable sources are located at relatively high redshift ($z>0.2$). In conjunction with the observation of the stochastic GW background from the whole population of MBHBs, the identification of individual MBHBs could provide new constraints on the populations of these objects and the relevant physical processes.

As a by-product of the analysis, we have also estimated the level of the GW stochastic background produced by the different models, finding good agreement with the estimates derived using merger tree realisations based on the Extended Press \& Schechter formalism considered in PaperI. Such agreement provides a further validation of the results of this paper and PaperI, it shows that we can now have in hand self-consistent predictions for stochastic and deterministic signals from the cosmic population of MBHBs, and suggests that EPS merger trees could provide a valuable approach to the studies of MBH evolution at low-to-medium redshift. 

As a final word of caution, we would like to stress that the results of this paper clearly suffer from considerable uncertainties determined by the still poor quantitative information about several parameters that control the models. The spread of the predictions of the expected events is therefore likely dominated by the lack of knowledge of the model parameters, rather than the differences between the assembly scenarios.

\end{document}